\documentclass[11pt,a4paper]{report}

\usepackage{amsmath}
\usepackage{latexsym}
\usepackage{amssymb}
\usepackage{amsfonts}
\usepackage{chapterbib}
\usepackage[dvips]{graphicx}
\usepackage{epic}
\usepackage{eepic}
\usepackage{psfrag}
\usepackage{epsfig}
\usepackage{german}

\begin{document}

\begin{center}
\mbox{}\\
{\Huge \textbf{Schr\"odingers Entdeckung der Wellenmechanik}\footnote{Vortrag anl\"asslich des Symposiums: SCHR\"ODINGER'S WAVEMECHANICS 75 YEARS AFTER (Universit\"at Z\"urich, 24-25 April 2001).}}\\
\vspace{1cm}
{\Large \textbf{Norbert Straumann}}\\
{\textbf{Institut f\"ur Theoretische Physik, Universit\"at Z\"urich, Winterthurerstrasse 180, 8057 Z\"urich, Switzerland}}\\
\end{center}

\section*{Einleitung}
In der ersten H\"alfte des Jahres 1926 hat Schr\"odinger im Alleingang die Wellenmechanik begr\"undet und in sechs Publikationen auch erstaunlich weit entwickelt [1,2]. Dieser sch\"opferische Ausbruch d\"urfte hinsichtlich Bedeutung und Intensit\"at nur ganz wenige Parallelen in der Geschichte der Naturwissenschaften haben.

In diesen Monaten hat Schr\"odinger die Physik in einer Weise ver\"andert und in neue Richtungen gewiesen wie das nur sehr wenigen verg\"onnt ist. Bereits um 1960 z\"ahlte man \"uber 100'000 wissenschaftliche Arbeiten, die ziemlich direkt auf Schr\"odinger aufbauen. W\"urde man allein abz\"ahlen, wie oft Autoren in diesem Auditorium das Wort \textit{Schr\"odingergleichung} in ihren Publikationen verwendet haben, d\"urfte man auf eine stattliche Zahl kommen.

Schr\"odingers Arbeiten sind nicht nur bedeutend, sondern auch in einem hinreissenden, eleganten Stil geschrieben. Ich m\"ochte versuchen, Ihnen etwas vom 
Vergn\"ugen weiterzugeben, das ich bei deren Lekt\"ure empfunden habe. Sein eigenes Erstaunen und die Lust an der Arbeit spiegeln sich auch im Briefwechsel dieser Monate. Auch davon m\"ochte ich Ihnen einige Kostproben weitergeben. (Bei dieser Gelegenheit ist es mir ein Anliegen, Karl von Meyenn f\"ur die Briefe zu danken, die er mir freundlicherweise \"ubergeben hat. Er plant, nach Abschluss des Pauli-Unternehmens, auch die Schr\"odinger-Korrespondenz herauszubringen [3].)

Schr\"odinger war f\"ur seine Pioniertat in jeder Hinsicht bestens ger\"ustet. Man bekommt den Eindruck, dass ihm -- von den Interpretationsfragen abgesehen -- das meiste regelrecht ``zufiel''. Das kommt z. B. in einem Brief vom 22. Feb. 1926 an Wilhelm Wien, den Herausgeber der 'Annalen' sehr sch\"on zum Ausdruck [4]. Schr\"odinger schreibt:\\*

\textit{"Die Zeit vergeht mir im Fluge. Jeder zweite oder dritte Tag bringt wieder eine kleine Neuigkeit }- es \textit{arbeitet, nicht ich, und dieses Es ist die herrliche klassische Mathematik und die Hilbert-Mathematik, das wundervolle Geb\"aude der Eigenwerttheorie. Die breiten alles so klar vor einem aus, dass 
man es nur zu nehmen braucht ohne M\"uhe und ohne Sorge, das Rechte stellt 
sich zu seiner Zeit, sobald man es braucht, ganz von selbst ein. Ich bin so 
froh, der schrecklichen Mechanik entronnen zu sein mit ihren Wirkungs- und 
Winkelvariablen und St\"orungstheorie, die ich nie recht verstanden habe. Jetzt ist alles} linear \textit{geworden, alles superponierbar, man rechnet so leicht und angenehm, wie in der alten Akustik. Und die St\"orungstheorie ist nicht 
komplizierter als die erzwungenen Schwingungen einer Saite.''}\\*

Da die Matrizenmechanik schon seit einigen Monaten vorlag, will ich aus diesem 
Brief auch Teile des Abschnitts mit dem Untertitel ``Verh\"altnis zu Heisenberg'' vorlesen. Man sieht daraus wie wichtig, ja produktiv, auch in der Forschung Vorlieben und Abneigungen sein k\"onnen. Ich zitiere:\\*

\textit{``Ich bin mit Geheimrat Sommerfeld von einer innerlich nahen Beziehung 
\"uberzeugt. Sie muss aber ziemlich tief liegen, denn} Weyl\textit{, der die 
Heisenbergsche Theorie sehr gr\"undlich studiert und selbst weiterentwickelt 
hat, und dem ich mein erstes Manuskript zu lesen gab, sagt, er weiss das 
Verbindungsglied nicht zu finden. Daraufhin hab' ich es aufgegeben, 
meinerseits zu suchen. Und das umso lieber, als mir der Matrizenkalk\"ul 
unertr\"aglich war lange bevor ich an meine Theorie auch nur entfernt dachte - 
ich will damit sagen, es ist nicht Vorliebe f\"ur das eigene Kind, die jetzt von nichts anderem wissen will.} Jetzt \textit{hoffe ich nat\"urlich fest, dass der Matrizenkalk\"ul, nach Aufsaugung seiner wertvollen Resultate durch die 
Eigenwerttheorie, wieder verschwinden wird. Ich glaube ehrlich gegen mich 
selbst zu sein, wenn ich sage, dass ich das nicht weniger heiss w\"unschen 
w\"urde, wenn die Eigenwerttheorie von Herrn Schultze herr\"uhrte. Denn mir 
schaudert vor dem blossen Gedanken, den Matrizenkalk\"ul sp\"ater einmal 
einem jungen Studenten als das wahre Wesen des Atoms vortragen zu m\"ussen.''}

\section*{Zur Vorgeschichte}

Bevor ich mich Schr\"odingers Serie zur Wellenmechanik zuwende, sind ein paar 
Bemerkungen zur Vorgeschichte wesentlich.

In seiner vierten Arbeit `` \"Uber das Verh\"altnis der Heisenberg-Born-Jordan- \\
schen Quantenmechanik zu der meinen'', schreibt Schr\"odinger in einer Fussnote: \textit{``Angeregt wurde meine Theorie durch L. de Broglie (...) und durch kurze aber unendlich weitblickende Bemerkungen A. Einsteins (...).''} Einsteins Beitrag zur Geburt der Wellenmechanik hat A. Pais in einem eigenen Kapitel (24) seines wunderbaren Einstein-Buches geschildert [5]. Ich muss mich hier auf das Allerwichtigste beschr\"anken.

Einstein erhielt von Langevin ein Exemplar der Dissertation von de Broglie und war davon beeindruckt. So schrieb er im Dezember 1924 an Lorentz [6]: \textit{``Ich glaube, das ist ein erster schwacher Strahl zur Erhellung dieses schlimmsten unserer physikalischen R\"atsel. Ich habe einiges gefunden, was f\"ur seine Konstruktion spricht.''}

Im Anschluss an Boses neuer Herleitung der Planckschen Strahlungsformel (basierend auf der Ununterscheidbarkeit der Photonen) schrieb Einstein drei Arbeiten \"uber Quantengase. Das heute bekannteste Resultat ist seine Herleitung der sog. Bose-Einstein-Kondensation, die seit ein paar Jahren wieder sehr aktuell geworden ist. Grundlegender ist aber Einsteins Untersuchung der Fluktuationen eines Quantengases [7], in enger Analogie zu seiner Arbeit von 1909 \"uber die Schwankungen der Strahlung [8]. Die Gleichartigkeit der beiden ber\"uhmten Schwankungsterme f\"uhrte ihn nun dazu, den Term, der bei der Strahlung mit Wellen verkn\"upft war, \textit{`` auch beim Gase in entsprechender Weise [zu] deuten, indem man dem Gase in passender Weise einen Strahlungsvorgang zuordnet und dessen Interferenz-Schwankungen berechnet. (...) Ich gehe n\"aher auf diese Deutung ein, weil ich glaube, dass es sich um mehr als eine blosse Analogie handelt.''} An dieser Stelle verweist Einstein auf de Broglies Dissertation, von der er sagt, dass sie \textit{``h\"ochste Aufmerksamkeit verdient''.}

Aus der Analyse der Schwankungen folgert also Einstein, dass die Existenz von Materiewellen  unausweichlich ist. In seinem Aufsatz Einsteins Beitrag zur Quantentheorie schrieb W. Pauli in diesem Zusammenhang [9]:\\*

\textit{``Der Autor erinnert sich, dass w\"ahrend einer Diskussion bei der 
Physikertagung in Innsbruck im Herbst 1924 Einstein die Suche nach 
Interferenz- und Beugungserscheinungen bei Molekularstrahlen vorschlug.''}\\*

Die letzte Arbeit von Schr\"odinger vor seiner Entdeckung der Wellenmechanik tr\"agt den Titel ``Zur Einsteinschen Gastheorie'' [10]. Ich kann darauf nicht n\"aher eingehen\footnote{Siehe A. Pais: p. 445-446.}, zitiere aber den folgenden Schl\"usselsatz:\\*

\textit{``Das heisst nichts anderes, als Ernst zu machen mit der de Broglie-
Einsteinschen Undulationstheorie der bewegten Korpuskeln (...).''}\\*

Am 23. Nov. 1925 gab Schr\"odinger in Z\"urich ein theoretisches Seminar \"uber de Broglies Dissertation. Felix Bloch hat vor vielen Jahren in einem Vortrag hier in Z\"urich davon berichtet und seine Erinnerungen auch in einem Artikel in ``Physics Today'' festgehalten [11]. Felix Bloch berichtet:\\*

\textit{``Debye casually remarked that he thought this way of talking was rather childish. As a student of Sommerfeld he had learned that, to deal properly 
with waves, one had to have a wave equation ... Just a few weeks later 
[Schr\"odinger] gave another talk in the colloquium which he started by saying: 'My colleague Debye suggested that one should have a wave equation; well, I 
have found one!' ''}\\*

Tats\"achlich stellte Schr\"odinger bereits Ende November (vielleicht auch etwas sp\"ater) zun\"achst eine station\"are relativistische Gleichung auf, die heute als Klein-Gordon-Gleichung bekannt ist. Diese l\"oste er anfangs Januar (1926) f\"ur das H-Atom. Dazwischen liegt ein geheimnisumwitterter Aufenthalt in Arosa, wo Schr\"odinger in der Villa Herwig mit einer Freundin aus Wien die Weihnachtsferien verbrachte, w\"ahrend seine Frau zu Hause blieb. In der ausgezeichneten Schr\"odinger-Biographie schreibt W. Moore dar\"uber [12]:\\*

\textit{``Like the dark lady who inspired Shakespeare's sonnets, the lady of Arosa may remain forever mysterious. We know that she was not Lotte or Irene. In 
all likelihood she was not Felicie; her husband had lost his fortune in the 
postwar inflation and had gone to Brasil, leaving her with an infant daughter. 
Whoever may have been his inspiration, the increase in Erwin's powers was 
dramatic, and he began a twelve-month period of sustained creative activity 
(...).''}\\*

Schr\"odinger war jedenfalls in Arosa, denn vom 27. Dez. datiert ein Brief an W. Wien aus dem hervorgeht [13], dass er am Eigenwertproblem f\"ur das H-Atom arbeitete, aber mit der radialen Differentialgleichung noch nicht ganz zurecht kam. Dies lag auch daran, dass er das von ihm damals benutzte Buch von L. Schlesinger \"uber Differentialgleichungen nicht mit nach Arosa genommen hatte. Der Brief an Wien enth\"alt den Seufzer: \textit{``Wenn ich nur mehr Mathematik k\"onnte.''}

Am 9. Januar kehrte Schr\"odinger nach Z\"urich zur\"uck. Auf die Frage von Dekan Schlaginhausen, ob er das Skifahren genossen habe, antwortete Schr\"odinger lakonisch, er sei durch ein paar Rechnungen davon abgehalten worden. In Z\"urich suchte er Hilfe bei H. Weyl und l\"oste innert Tagen das Wasserstoffproblem.

\section*{Die relativistische Gleichung}

Bevor ich zur ersten Publikation von Schr\"odinger \"uber Wellenmechanik komme, m\"ochte ich noch erkl\"aren, wie er die station\"are Klein-Gordon-Gleichung fand. Dar\"uber geben gewisse Forschungsnotizen Auskunft (von Moore mit (N1) und (N2) bezeichnet).

Zum einfacheren Verst\"andnis gebe ich zuerst die \underline{nichtrelativistische} Version seiner Argumentation. Als Ausgangspunkt benutzt Schr\"odinger die station\"are Wellengleichung
\begin{equation}
(\Delta+k^2)\psi=0.
\end{equation}
Zur Bestimmung der Wellenzahl verwendet er die erste der Einstein-de Broglie-Beziehungen
\begin{equation}
\mathbf{p}=\hbar \mathbf{k}, \; \; E=\hbar \omega,
\end{equation}
in der mechanischen Gleichung
\begin{equation}
E-V=\frac{1}{2m}{\mathbf{p}}^2.
\end{equation}
In die Wellengleichung (1) eingesetzt gibt dies bereits die station\"are Schr\"odinger-Gleichung f\"ur das Einelektronenproblem:
\begin{equation}
(-\frac{\hbar^2}{2m}\Delta+V)\psi=E\psi.
\end{equation}
Die relativistische Erweiterung ist unmittelbar: Anstelle von (3) haben wir
\begin{equation*}
E-V=E_{kin}=\sqrt{c^2\mathbf{p}^2+(mc^2)^2},
\end{equation*}
also
\begin{equation}
\mathbf{p}^2=\hbar^2 \mathbf{k}^2=\frac{1}{c^2}[(E-V)^2-(mc^2)^2].
\end{equation}
In (1) eingesetzt ergibt sich jetzt die station\"are Klein-Gordon-Gleichung
\begin{equation}
\Big[\Delta+\Big(\frac{E-V}{\hbar c}\Big)^2-\Big(\frac{mc}{\hbar}\Big)^2\Big]\psi=0.
\end{equation}

\section*{Die erste Mitteilung}

Nun beginne ich mit der Besprechung der sechs Schr\"odingerschen Arbeiten [1] zur 
Wellenmechanik (mit W1-W6 abgek\"urzt). Dabei werde ich die Gewichte aus guten Gr\"unden sehr ungleichm\"assig verteilen. Auf seine zweite Mitteilung m\"ochte ich besonders detailliert eingehen.

Angeblich gab es eine 1. Version von Schr\"odingers erster Mitteilung, welche die relativistische Gleichung und deren Anwendung auf das H-Atom zum Gegenstand hatte. Da die Spin-0 Gleichung bekanntlich die falsche Feinstruktur liefert, soll Schr\"odinger diese aber zur\"uckgezogen haben. Freilich sind dazu, soweit mir bekannt ist, keine eindeutigen Dokumente gefunden worden\footnote{ Viel sp\"ater best\"atigte aber S. die Existenz dieses Manuskripts. In einem Brief an 
W. Yourgrau vom Jan. 1956 schrieb er [14]: \textit{"My paper in which this is shown has ... never been published, it was withdrawn by me and replaced by the non-relativistic treatment.''}}.

Sie w\"urden wahrscheinlich erwarten, dass Schr\"odinger mit einer Begr\"undung seiner station\"aren Wellengleichung beginnt, die etwa so verl\"auft wie ich das eben ausgef\"uhrt habe. Stattdessen basiert er sie auf einer Hypothese, die ich obskur finde, und es ist mir nicht klar was dahintersteckt. Zun\"achst beginnt er triumphierend:\\*

\textit{``\S1. In dieser Mitteilung m\"ochte ich zun\"achst an dem einfachsten Fall des (nichtrelativistischen und ungest\"orten) Wasserstoffatoms zeigen, dass die \"ubliche Quantisierungsvorschrift sich durch eine andere Forderung ersetzen l\"asst, in der kein Wort von ``ganzen Zahlen'' mehr vorkommt. Vielmehr ergibt sich die Ganzzahligkeit auf dieselbe nat\"urliche Art, wie etwa die 
Ganzzahligkeit der Knotenzahl einer schwingenden Saite. Die neue 
Auffassung ist verallgemeinerungsf\"ahig und r\"uhrt, wie ich glaube, sehr tief an das wahre Wesen der Quantenvorschriften.''}\\*

Danach erinnert er an die station\"are Hamilton-Jacobi-Gleichung. Diese schreibt er auf $\psi=e^{S/K}$ um, wobei $K$ sp\"ater gleich unserem heutigen $\hbar$ gesetzt wird. Speziell f\"ur das 1-Elektronenproblem im Coulombfeld erh\"alt Schr\"odinger
\begin{equation}
(\mathbf{\nabla}\psi)^2-\frac{2m}{\hbar^2}\Big(E+\frac{e^2}{r}\Big)\psi^2=0.
\end{equation}
Nun postuliert er, dass die linke Seite als Dichte eines Variationsprinzips genommen werden soll:
\begin{equation}
\delta\int \Big[(\mathbf{\nabla}\psi)^2-\frac{2m}{\hbar^2}\Big(E+\frac{e^2}{r}\Big)\psi^2 \Big]d^3x=0.
\end{equation}
Die zugeh\"orige Euler-Lagrange-Gleichung kann sofort abgelesen werden und stimmt mit (4) f\"ur $V= -e^2/r$ \"uberein. In (W2) weist Schr\"odinger selber auf die Unverst\"andlichkeit dieser Begr\"undung hin, die man bestensfalls als sehr formal charakterisieren muss.

Die Separation von (4) f\"ur ein zentralsymmetrisches Potential ist f\"ur ihn nat\"urlich kein Problem und es bleibt als Hauptaufgabe die \"uberall endlichen  L\"osungen der radialen Gleichung f\"ur das H-Atom zu bestimmen. Er beginnt mit der Bemerkung, dass letzere in der komplexen $r$-Ebene Singularit\"aten bei $r=0$ und $r=\infty$ hat und charakterisiert die Natur dieser Singularit\"aten. An dieser Stelle bemerkt er in einer Fussnote: \textit{``F\"ur die Anleitung zur Behandlung der Gleichung (...) bin ich Hermann Weyl zu gr\"osstem Dank verpflichtet''} (und er verweist im \"ubrigen auf das bereits erw\"ahnte Buch von L. Schlesinger). Bei der anschliessenden Durchf\"uhrung kommt es einem tats\"achlich so vor, als ob ihm Weyl \"uber die Schultern blickt. Schr\"odingers Behandlung ist wesentlich strenger als in den meisten Lehrb\"uchern. Hier zeigen sich Liebe und Talent f\"ur die Mathematik.

Im letzten Abschnitt macht er ein paar vorl\"aufige interpretatorische Bemerkungen und erw\"ahnt nebenbei das Resultat f\"ur die relativistische Gleichung.

\section*{Erste Reaktionen}

Nach dieser v\"ollig neuartigen Erkl\"arung des Wasserstoffspektrums horchte die Welt auf. Die Arbeit wurde am 27. Jan. (1926) \"uber W. Wien bei den Annalen eingereicht, mit der Bitte, diese auch Sommerfeld zu zeigen. Zwei Tage sp\"ater berichtete Schr\"odinger in einem Brief an Sommerfeld [15] \"uber weitere Resultate zum linearen Oszillator und zum Rotator, und k\"undete anstehende Untersuchungen, u.a. zum Stark-Effekt an.

Die Antwort von Sommerfeld vom 3. Februar beginnt mit [16]:\\*

\textit{``Das ist ja furchtbar interessant was Sie schreiben, in Abhandlung und Brief. 
(...) Ich war gerade dabei, f\"ur Vortr\"age in London (diesen M\"arz) ein Konzept zu machen, das auf der fr\"uheren Tonart blies. Da traf, wie ein Donnerschlag, Ihr Manuskript ein.''}\\*

Etwas sp\"ater sagt Sommerfeld: \textit{``Dass [das Ergebnis \"uber den Rotator] mit dem uralten $n(n+1)$ der Kugelfunktionen zusammenh\"angen soll, geht \"uber jede Hutschnur.''} Ich zitiere noch ein paar weitere S\"atze aus dem interesssanten Brief:\\*

\textit{``Nat\"urlich \"ubersehe ich mathematisch noch gar nicht, wie das alles 
zusammenh\"angt, aber ich bin \"uberzeugt, dass etwas ganz neues daraus 
werden wird, was die Widerspr\"uche beseitigen kann, die uns jetzt sekieren. 
(...) Eigent\"umlich ist die Verschiedenheit der Ausgangspunkte bei Gleichheit der Resultate zwischen Ihnen und Heisenberg''.}\\*

Noch am gleichen Tag schrieb Sommerfeld in einem Brief an Pauli [17]:\\*

\textit{``Hier ist ein Manuskript von Schr\"odinger f\"ur die Annalen eingelaufen. Schr. scheint ganz dieselben Resultate zu finden, wie Heisenberg und Sie, aber auf einem ganz anderen, total verr\"uckten, Wege, keine Matrixalgebra, sonder Randwertaufgaben. Sicher wird aus alle dem bald in irgend einer Form etwas 
Vern\"unftiges und Definitives entstehen.''}\\*

Kurz nachdem Schr\"odingers erste Mitteilung erschienen war, schrieb Pauli am 12. April einen langen, sehr bemerkenswerten Brief an Jordan [18], in welchem er ihm die Resultate von einigen \"Uberlegungen mitteilt, \textit{``die ich im Zusammenhang mit der in den Annalen der Physik erschienenen Arbeit \"uber 'Quantisierung als Eigenwertproblem' von Schr\"odinger angestellt habe''.} Pauli bemerkt dazu gleich: \textit{``Ich glaube, dass diese Arbeit mit zu dem Bedeutendsten z\"ahlt, was in letzer Zeit geschrieben wurde. Lesen Sie sie sorgf\"altig und mit Andacht''.}\\*

Im n\"achsten Satz wird der Hauptinhalt des Briefes angedeutet:\\*

\textit{``Nat\"urlich habe ich mich gleich gefragt, wie seine Resultate mit denen der G\"ottinger Mechanik zusammenh\"angen. Diesen Zusammenhang glaube ich 
jetzt vollst\"andig verstanden zu haben.''}\\*

Sie sehen, wie damals Neuigkeiten die Runde machten. Auf weitere Reaktionen von bedeutenden Theoretikern werde ich zur\"uckkommen.

\section*{Die zweite Mitteilung}

Bereits vier Wochen nach der 1. Mitteilung reichte Schr\"odinger seine zweite bei den Annalen ein. Hier geht es ihm mit seinen Worten darum \textit{``den \underline{allgemeinen} Zusammenhang n\"aher zu beleuchten, welcher zwischen der Hamiltonschen partiellen Differentialgleichung eines mechanischen Problems und der 'zugeh\"origen' Wellengleichung (...) besteht.''}

Ein wesentlicher Teil dieser l\"angeren Arbeit von $\sim \,$40 Seiten ist dabei der Hamilton-Jacobischen Theorie, insbesondere der Hamiltonschen Analogie zwischen Mechanik und Optik gewidmet (\S1). Als Rechtfertigung dieser ausf\"uhrlichen Exposition sagt Schr\"odinger:\\*

\textit{``Leider ist dieser kraftvolle und folgenschwere Ideenkreis Hamiltons in den meisten modernen Wiedergaben seines sch\"onen anschaulichen Gewandes 
als eines \"uberfl\"ussigen Beiwerks beraubt worden zugunsten einer mehr 
farblosen Darstellung der analytischen Zusammenh\"ange.''}\\*

Und er weist in einer Fussnote auf F. Klein hin, der die grosse Bedeutung der 
Hamiltonschen optischen Abhandlungen schon viel fr\"uher betont hatte, aber entt\"auscht zur Kenntnis nehmen musste, dass seine Bem\"uhungen nicht viel gefruchtet hatten.

Schr\"odingers Darstellung dieser bekannten Zusammenh\"ange ist in einem wundersch\"onen Stil geschrieben. Seine Art von wissenschaftlicher Prosa ist l\"angst aus den physikalischen Zeitschriften verschwunden. Er beschreibt die mathematischen Sachverhalte g\"anzlich informell, aber trotzdem pr\"azise, - nat\"urlich ohne all die Pedanterie die man heute als n\"otig erachtet. Eine kurze Wiedergabe des Inhalts ist an dieser Stelle n\"otig, um Schr\"odingers  Begr\"undung seiner Wellengleichung, auch f\"ur Mehrelektronensysteme, zu verstehen. Diese kann man \"ubrigens (in etwas gewandelter Form) auch f\"ur den heutigen Unterricht noch sehr empfehlen.

Der Ausgangspunkt von \S1 ist die Hamilton-Jacobi (HJ)-Gleichung f\"ur ein autonomes Hamiltonsches System. F\"ur ein solches kann die Wirkungsfunktion $S(q^i,t)$ bekanntlich in der Form
\begin{equation}
S(q^i,t)=W(q^i)-Et
\end{equation}
gew\"ahlt werden und $W(q^i)$ erf\"ullt die \underline{verk\"urzte} HJ-Gleichung:
\begin{equation}
H\Big(\frac{\partial W}{\partial q},q\Big)=E.
\end{equation}
Schr\"odinger spezialisiert f\"ur alles Weitere die Hamiltonfunktion $H(p,q)$ auf
\begin{equation}
H(p,q)=T(p,q)+V(q),
\end{equation}
wobei die kinetische Energie $T$ eine quadratische Form in den kanonischen Impulsen ist
\begin{equation}
T(p,q)=\frac{1}{2}\sum_{i,j}g^{ij}(q)p_i p_j.
\end{equation}
Im Lagrange-Formulismus ist
\begin{equation}
T(q,\dot{q})=\frac{1}{2}\sum_{i,j}g_{ij}(q)\dot{q}^i \dot{q}^j.
\end{equation}
Die $g_{ij}(q)$ definieren, wie Schr\"odinger mit einem Hinweis auf H. Hertz sehr betont, eine nichteuklidische Massbestimmung (eine Riemannsche Metrik)
\begin{equation}
ds^2=g_{ij}(q)dq^i dq^j
\end{equation}
($g^{ij}(q)$ in (12) ist die inverse Matrix). Differentialoperatoren im $q$-Raum --etwa der Laplace-Operator-- sind mit dieser Metrik zu bilden.\\*
In Schr\"odingers Worten: \textit{``Wir setzen fest, dass im folgenden alle geometrischen Aussagen im $q$-Raum in diesem nichteuklidischen Sinn zu verstehen sind.''}

Im Anschluss daran schreibt Schr\"odinger die verk\"urzte HJ-Gleichung f\"ur (11) so (seine Gl. 1''):
\begin{equation}
\|\nabla W \|^2=E-V.
\end{equation}
Die Fl\"achen {$S=const$} interpretiert er als \textit{``System der Wellenfl\"achen einer fortschreitenden, aber station\"aren Wellenbewegung im $q$-Raum''} und bestimmt deren \underline{Phasengeschwindigkeit}.

Bevor wir dies nachvollziehen, sollte ich an den folgenden Sachverhalt der HJ-
Theorie erinnern. Zu einer L\"osung $W(q)$ der partiellen Differentialgleichung (10) ist es nat\"urlich, die folgende gew\"ohnliche Differentialgleichung im $q$-Raum (Konfigurationsraum) zu betrachten:
\begin{equation}
\dot{q}(t)=\frac{\partial H}{\partial p}\Big(\frac{\partial W}{\partial q}(q(t)),q(t)\Big).
\end{equation}
Jede L\"osung dieser Gleichung beschreibt eine mechanische Bahn im Konfigurationsraum, ist also eine L\"osung der Euler-Lagrange-Gleichung. Tats\"achlich l\"asst sich zeigen, dass als Folge dieser Gleichung sowie der HJ-Gleichung (10), $q(t)$ und
\begin{equation}
p(t):=\frac{\partial W}{\partial q}(q(t))
\end{equation}
dem Paar der Hamiltonschen Gleichungen gen\"ugen.
F\"ur die Klasse (11) lautet (16)
\begin{equation}
\dot{q}(t)=\nabla W(q(t)) \;\;\;\; \Big(\dot{q}^i(t)=g^{ij}\frac{\partial W}{\partial q^j}\Big),
\end{equation}
folglich sind die mechanischen Bahnen senkrecht zur Schar ($W=const$). (In der Mathematik l\"auft dies unter dem Begriff der Charakteristiken einer nichtlinearen partiellen Differentialgleichung  1. Ordnung.)

Nun wollen wir mit Schr\"odinger die \underline{Fl\"achen konstanter Phase} {$S(q,t) = const$} betrachten. F\"ur $W$ bedeutet dies $W=const+Et$. Bei gegenbener Konstante betrachten wir die Fl\"achen zur Zeit $t$ und $t+dt$ (s. Abb. 1), sowie eine mechanische Bahn $q(t)$, die wir auch mit der Bogenl\"ange $s$ zur Metrik (14) parametrisieren k\"onnen.

\vspace{1cm}

\begin{figure}[h]
\begin{center}
\epsfig{file=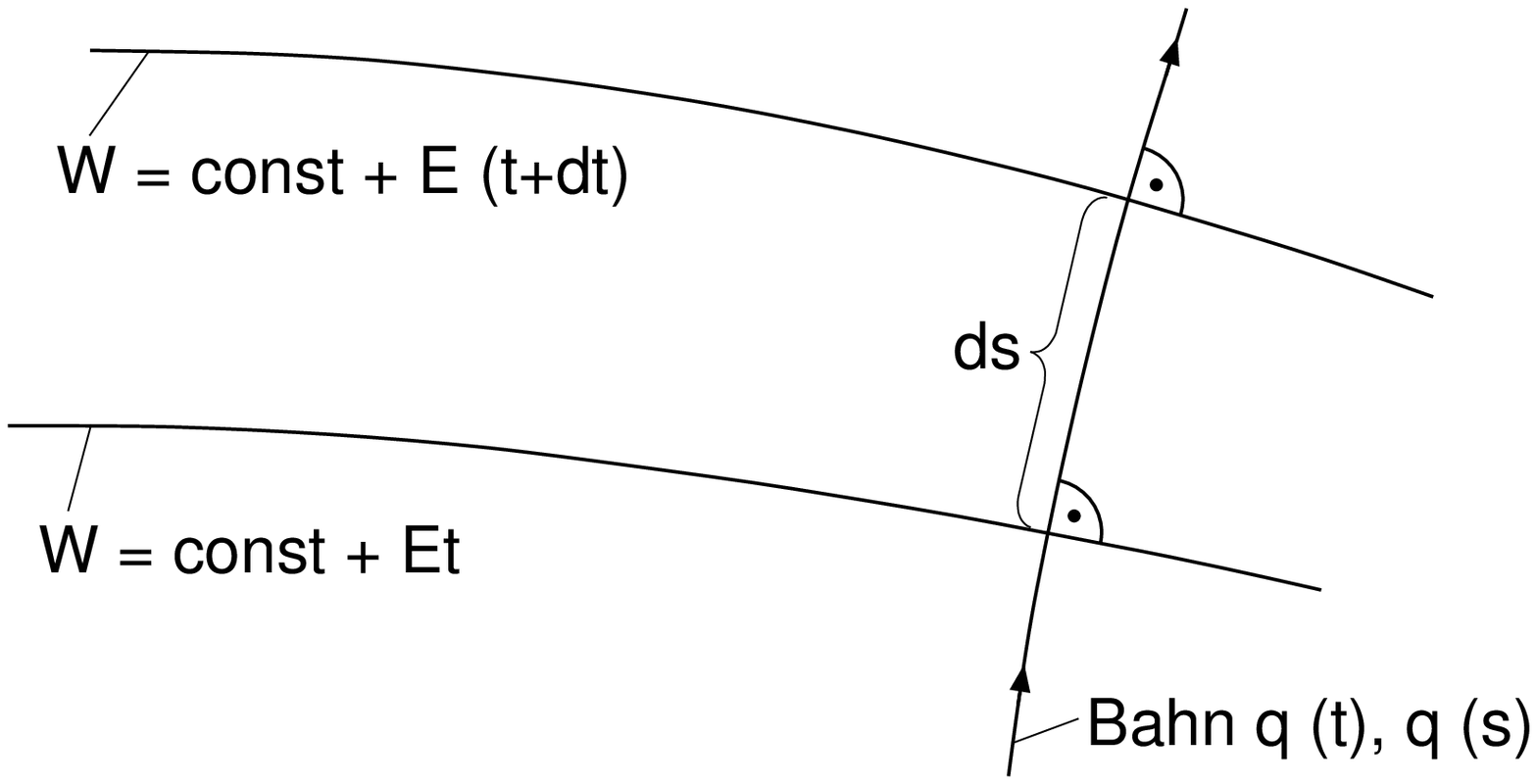,width=12cm,angle=0}
\vspace{1cm}
\hspace{5cm} \textbf{Abb. 1}: Fl\"achen konstanter Phase und mechanische Bahnen als Orthogonaltrajektorien.
\end{center}
\end{figure}

\vspace{1cm}

L\"angs der Bahn gilt einerseits nach der Kettenregel
\begin{equation*}
\frac{d W(q(s))}{d s}=E\frac{d t}{d s}
\end{equation*}
und andererseits finden wir, da $\dot{q}(t)$ und $\nabla W$ parallel sind,
\begin{equation*}
\frac{d W(q(s))}{d s}=<dW,\frac{dq}{ds}>=\Big(\nabla W,\frac{dq}{ds}\Big)=\| \nabla W \|
\end{equation*}
(beim letzten Gleichheitszeichen wurde $\|\frac{dq}{ds}\|=1$ verwendet). Also ist die Phasengeschwindigkeit
\begin{equation}
u:=\frac{d s}{d t}=\frac{E}{\| \nabla W \|}=\frac{E}{\sqrt{2(E-V)}}.
\end{equation}
(Dies ist bei Schr\"odinger die Gl. (6).)
Damit folgt nun sofort, und dies wird von Schr\"odinger herausgestrichen, dass das \underline{Hamiltonsche Prinzip in der Maupertuis-} \\ \underline{schen Form und das Fermatsche Prinzip zusammenfallen}.

Ich erinnere kurz an ersteres. Betrachten wir eine mechanische Bahn $q(t)$ auf der zugeh\"origen Energiefl\"ache im Phasenraum (s. Abb. 2) und betten diese in die Schar von Phasentrajektorien zu gleicher Energie und gleichen Anfangs- und Endpunkten ein, dann ist die reduzierte Wirkung $\int p \, dq$ f\"ur die tats\"achliche Bahn $q(t)$ \underline{station\"ar}. Es gilt also das Variationsprinzip [19]
\begin{equation}
\delta \int p \, dq=0.
\end{equation}

\vspace{1cm}

\begin{figure}[h]
\begin{center}
\epsfig{file=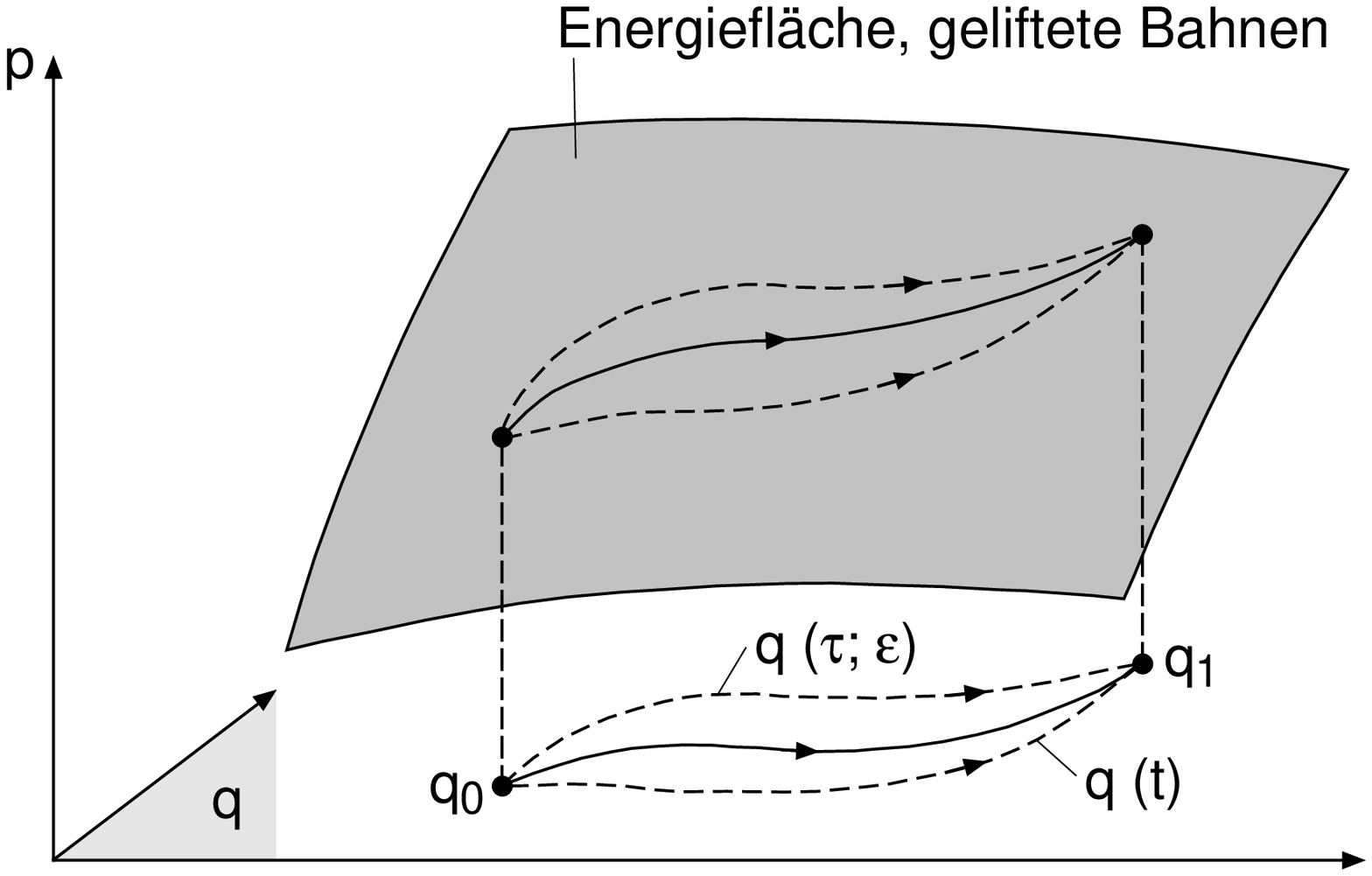,width=12cm,angle=0}
\vspace{1cm}
\hspace{5cm} \textbf{Abb. 2}: Zum Prinzip von Maupertuis (Euler-Lagrange-Jacobi).
\end{center}
\end{figure}

\vspace{1cm}

Da f\"ur die Konkurrenzschar der Zeitparameter $\tau$ so gew\"ahlt werden muss, dass
\begin{equation*}
H\Big(\frac{\partial L}{\partial \dot{q}}(\tau),q(\tau)\Big)=E
\end{equation*}
gilt, haben wir mit
\begin{equation}
\frac{\partial L}{\partial \dot{q}} \, \dot{q}=2T=2(E-V)\;\;\; \mbox{und} \;\; \; 2T=(\frac{ds}{d \tau})^2
\end{equation}
die Beziehung
\begin{equation}
d \tau=\frac{ds}{\sqrt{2(E-V)}}.
\end{equation}
Damit k\"onnen wir die reduzierte Wirkung so umschreiben:
\begin{equation}
\int p \, dq=\int \frac{\partial L}{\partial \dot{q}} \, \dot{q} \, d \tau=\int 2(E-V) d \tau = \int \sqrt{2(E-V)} ds=E \int \frac{ds}{u},
\end{equation}
d.h. \underline{die reduzierte Wirkung ist proportional zur Wirkung im Fermatprinzip}.\\
Dies ist ein besonders sch\"oner Ausdruck der mechanisch-optischen Analogie.

Den Abschnitt 1 beschliesst Schr\"odinger mit verschiedenen zusammenfassenden 
und weiterf\"uhrenden Bemerkungen. Zun\"achst betont er mit Recht (und das muss man gelegentlich auch heute noch wiederholen):\\*

\textit{``Trotzdem in den vorstehenden \"Uberlegungen von Wellenfl\"achen, 
Fortpflanzungsgeschwindigkeit, Huygensschem Prinzip die Rede ist, hat 
man dieselben doch eigentlich nicht als eine Analogie der Mechanik mit der 
\underline{Wellen}optik, sondern mit der \underline{geometrischen} Optik anzusehen.''}\\*

Etwas sp\"ater erg\"anzt er:\\*

\textit{``..Und das System der $S$-Fl\"achen, als Wellenfl\"achen aufgefasst, steht zur mechanischen Bewegung vorerst in einer etwas loseren Beziehung insofern, als der Bildpunkt des mechanischen Systems auf dem Strahl keineswegs etwa 
mit der Wellengeschwindigkeit u fortr\"uckt, sondern im Gegenteil, seine 
Geschwindigkeit ist (bei konstantem E) proportional zu $1/u$.''}\\*

Sie ergibt sich ja direkt als
\begin{equation}
v=\frac{ds}{dt}=\sqrt{2T}=\sqrt{2(E-V)}.
\end{equation}
Nachdem Schr\"odinger nochmals wiederholt, dass es sich bis jetzt ausschliesslich um \underline{geometrische} Optik handelte, sieht er die Parallele mit der Mechanik nicht bloss als \textit{``erfreuliches Anschauungsmittel''}. Denn, so sagt er, \textit{``schon der erste Versuch einer wellentheoretischen Ausgestaltung f\"uhrt auf so frappante Dinge, dass ein ganz anderer Verdacht aufsteigt: wir wissen doch heute, dass unsere klassische Mechanik bei sehr kleinen Bahndimensionen und sehr starken Bahnkr\"ummungen versagt. Vielleicht ist dieses Versagen eine volle Analogie zum Versagen der geometrischen Optik''.} Dar\"uber l\"asst er sich noch weiter aus und beschliesst die Betrachtung mit der Aussage:\\*

\textit{``Dann gilt es, eine ``undulatorische Mechanik'' zu suchen und der 
n\"achstliegende Weg dazu ist wohl die wellentheoretische Ausgestaltung des 
Hamiltonschen Bildes.''}\\*

Dies wird nun in \S2, unter dem Titel 'geometrische' und 'undulatorische' Mechanik, breit ausgef\"uhrt. Ich muss diese Diskussion stark k\"urzen.

In einer wellenmechanischen Ausstattung gibt es die Begriffe Frequenz und 
Wellenl\"ange. Nach Einstein und de Broglie gilt
\begin{equation}
\nu=\frac{E}{h}
\end{equation}
und f\"ur die Wellenl\"ange $\lambda$ dr\"angt sich nach (19) und (25)
\begin{equation}
\lambda=\frac{u}{\nu}=\frac{h}{\sqrt{2(E-V)}}
\end{equation}
auf. Die Wellenzahl $k=\omega/u$ ist dann
\begin{equation}
k=\frac{1}{\hbar}\sqrt{2(E-V)}.
\end{equation}
Zusammen mit (25) folgt das \underline{Dispersionsgesetz}
\begin{equation}
\omega=\frac{1}{\hbar}\Big[\frac{1}{2}(\hbar k)^2+V\Big],
\end{equation}
von dem Schr\"odinger sagt, dass es grosses Interesse bietet. Als Hauptpunkt ergibt sich, dass die zugeh\"origen Gruppengeschwindigkeit $d\omega/dk$ mit der mechanischen Geschwindigkeit \"ubereinstimmt:
\begin{equation}
\frac{d \omega}{dk}=\hbar k=\sqrt{2(E-V)}=v.
\end{equation}
Nach etwas l\"anglichen Betrachtungen sagt Schr\"odinger schliesslich: \textit{``Was ich mit grosser Bestimmtheit vermute ist folgendes:\\*
Das wirkliche mechanische Geschehen wird in zutreffender Weise erfasst oder abgebildet durch die \underline{Wellenvorg\"ange} im $q$-Raum.''}

Erneut schliesst er weitausgreifende \"Uberlegungen an, und stellt sich schliess-\\*lich die eigentliche Aufgabe:\\*

\textit{``Man muss statt von den Grundgleichungen der Mechanik von einer Wellengleichung f\"ur den $q$-Raum ausgehen und die Mannigfaltigkeit der nach ihr m\"oglichen Vorg\"ange betrachten.''}\\*

Aus Gr\"unden der Einfachheit machte er folgenden Vorschlag: Die \"ubliche station\"are Wellengleichung im $q$-Raum ist
\begin{equation}
(\Delta_g+k^2)\psi=0,
\end{equation}
wo $\Delta_g$ der Laplace-Operator zur Metrik (14) ist. Setzt man hier den Ausdruck (27) f\"ur $k$ ein, so kommt
\begin{equation}
\Big(-\frac{\hbar^2}{2}\Delta_g+V \Big)\psi=E\psi.
\end{equation}
Im Spezialfall eines Teilchens der Masse $m$ ist $\Delta_g=\frac{1}{2m}\Delta$, wo $\Delta$ der Laplace-Operator im euklidischen $\mathbf{R^3}$ ist, und (31) reduziert sich auf (4).

Mit der Gl. (31) waren nun auch die Grundlage f\"ur die Behandlung von mehreren Teilchen, freilich noch ohne Spin, gegeben. Von der Schr\"odinger-Gleichung (30) sagt der Autor:\\*

\textit{``Unser Ansatz ist wieder von dem Bestreben nach Einfachheit diktiert, doch halte ich diesfalls eine Irreleitung nicht f\"ur ausgeschlossen.''}\\*

Sehr ermutigend ist aber, dass \textit{``in allen F\"allen der klassischen Dynamik, die ich bisher untersuchte habe, die Gl. (31) die Quantenbedingungen in sich tr\"agt.''}

Schr\"odinger beendet diesen wichtigsten Abschnitt seiner Arbeit mit Anmerkungen zur Matrizenmechanik.\\*

\textit{``Ich m\"ochte an dieser Stelle die Tatsache nicht mit Stillschweigen \"ubergehen, dass gegenw\"artig von seiten Heisenbergs, Borns, Jordans und einiger anderer hervorragender Forscher ein Versuch zur Beseitigung der Quantenschwierigkeit im Gange ist, der schon auf so beachtenswerte Erfolge hinzuweisen hat, dass es schwer wird, daran zu zweifeln, er enthalte jedenfalls einen Teil der Wahrheit. In der} Tendenz \textit{steht der Heisenbergsche Versuch dem vorliegenden ausserordentlich nahe, davon haben wir schon oben gesprochen. In der Methode ist er so toto genere verschieden, dass es mir bisher nicht gelungen ist, das Verbindungsglied zu finden. Ich hege die ganz bestimmte Hoffnung, dass diese beiden Vorst\"osse einander erg\"anzen werden, indem der eine weiterhilft, wo der andere versagt. Die St\"arke des Heisenbergschen Programms liegt darin, dass es die Linienintensit\"aten zu geben verspricht, eine Frage, von der wir uns hier bisher ganz ferngehalten haben.''}\\*

Die weiteren Abschnitte von Schr\"odingers 2. Mitteilung sind dem harmonischen 
Oszillator und dem Rotator als Anwendungsbeispiele gewidmet. Ferner k\"undet er st\"orungstheoretische Untersuchungen, insbesondere zum Starkeffekt 1. Ordnung an.

Seine vorl\"aufigen Bemerkungen zur Interpretation der Wellenmechanik lasse ich vorderhand weg. Schr\"odinger war mit der Ausgestaltung seiner Theorie derart intensiv besch\"aftigt, dass er die damit zusammenh\"angenden Fragen etwas in den Hintergrund schob. Er hielt aber an einem \underline{klassischen} Wellenbild fest. Die damit verbundene Schwierigkeit, n\"amlich dass seine Wellenfunktion auf einem mehrdimensionalen Konfigurationsraum lebt, war ihm freilich durchaus bewusst. Ich werde auf die Interpretationsfragen am Schluss meines Vortrags zur\"uckkommen.

\section*{Der stetige \"Ubergang von der Mikro- zur Makromechanik}

Diesen Titel tr\"agt eine kurze Arbeit [1], die Schr\"odinger in den ``Naturwissenschaften'' publizierte. Darin zeigt er f\"ur den harmonischen Oszillator, \textit{``dass eine \underline{Gruppe} von Eigenschwingungen hoher Ordnungszahl $n$ ('Quantenzahl') und relativ kleinen Ordnungszahldifferenzen einen 'Massenpunkt' darzustellen vermag, welcher die nach der gew\"ohnlichen Mechanik zu erwartenden 'Bewegung' ausf\"uhrt (...)''.} Diese einfache Untersuchung beschliesst er mit der etwas erstaunlichen Bemerkung:\\*

\textit{``Es l\"asst sich mit Bestimmtheit voraussehen, dass man auf ganz \"ahnliche Weise auch die Wellengruppen konstruieren kann, welche auf hochquantigen 
Keplerellipsen umlaufen und das undulationsmechanische Bild des 
Wasserstoffelektrons sind; nur sind da die rechentechnischen Schwierigkeiten 
gr\"osser als in dem hier behandelten, ganz besonders einfachen Schulbeispiel.''}\\*

Damit waren aber nicht alle einverstanden. Hier ist vor allem H.A. Lorentz zu erw\"ahnen, der Schr\"odinger sofort zeigte, dass er irrte. Bereits in einem ersten von zwei langen Briefen, noch vor Schr\"odingers Note (W3), schreibt Lorentz unter Punkt 4 [20]:\\*

\textit{``Wenn wir uns dazu entschliessen, das Elektron sozusagen ganz aufzul\"osen 
und durch ein Wellensystem zu ersetzen, so hat das einen Nachteil und einen 
Vorteil.
Der Nachteil, und zwar ein schwerwiegender, ist dieser: was wir von 
dem Elektron des Wasserstoff-Atoms annehmen, m\"ussen wir wohl auch von 
allen Elektronen in allen Atomen voraussetzen; wir m\"ussen sie alle durch 
Wellensysteme ersetzen. Wie soll ich dann aber die Erscheinungen der Photo-
Elektrizit\"at und das Entweichen von Elektronen aus erhitzten Metallen 
verstehen? Hier kommen die Teilchen ganz nett und unversehrt zum 
Vorschein; wie haben sie sich wieder, einmal aufgel\"ost, zusammenballen 
k\"onnen?''}\\*

In einen zweiter Brief schreibt Lorentz [21]:\\*

\textit{``Mit der Zusendung Ihrer Note 'Der stetige \"Ubergang von der Mikro- zur Makromechanik' haben Sie mir eine grosse Freude gemacht, und als ich sie 
gelesen hatte, war mein erster Gedanke: mit einer Theorie, die einen Einwand 
in so \"uberraschender und sch\"oner Weise widerlegt, muss man schon auf dem 
rechten Wege sein. Leider hat sich meine Freude alsbald wieder getr\"ubt; ich 
kann n\"amlich nicht einsehen, wie Sie z.B. im Falle des Wasserstoffatoms 
Wellenpakete konstruieren k\"onnen, die (ich denke jetzt an die} sehr hohen 
\textit{Bohrschen Bahnen) sich wie das Elektron bewegen. Die dazu erforderlichen} kurzen \textit{Wellen stehen nicht zu Ihrer Verf\"ugung. Ich habe diesen Punkt schon 
in meinem ersten Briefe ber\"uhrt, und m\"ochte jetzt etwas n\"aher darauf 
eingehen.'' }\\*

Es folgt eine 12seitige Rechnung des damals 73j\"ahrigen Lorentz, mit dem Ergebnis, dass dies f\"ur das H-Atom tats\"achlich nicht m\"oglich ist.

Die Briefe von H.A. Lorentz sind auch dar\"uber hinaus sehr beeindruckend. Kein anderer Kollege ist derart detailliert auf die verbleibenden Schwierigkeiten eingegangen. Diesen Briefen m\"usste ich mehr Zeit widmen k\"onnen.

\section*{Wellenmechanik und Matrizenmechanik}
Aus der n\"achsten Arbeit [1] ``\"Uber das Verh\"altnis der Heisenberg-Born-Jordanschen Quantenmechanik zu der meinen'' zitiere ich zun\"achst ein paar S\"atze aus der Einleitung, die neben der Problemstellung auch einen Eindruck von Schr\"odingers Stil geben:\\*

\textit{``Bei der ausserordentlichen Verschiedenheit der Ausgangspunkte und 
Vorstellungkreise der Heisenbergschen Quantenmechanik einerseits und der 
neulich hier in ihren Grundz\"ugen dargelegten und als 'undulatorische' und 
'physikalische' Mechanik bezeichneten Theorie andererseits, ist es recht 
seltsam, dass diese beiden neuen Quantentheorien hinsichtlich der bisher 
bekannt gewordenen speziellen Ergebnisse} miteinander \textit{auch dort 
\"ubereinstimmen, wo sie von der alten Quantentheorie abweichen. Ich nenne 
vor allem die eigent\"umliche 'Halbzahligkeit' beim Oszillator und beim 
Rotator. Das ist wirklich sehr merkw\"urdig, denn Ausgangspunkt, 
Vorstellungen, Methode, der ganze mathematische Apparat scheinen in der 
Tat grundverschieden. Vor allem aber scheint das Abgehen von der 
klassischen Mechanik in den beiden Theorien geradezu in diametral 
entgegengesetzter Richtung zu erfolgen. Bei Heisenberg werden die 
klassischen kontinuierlichen Variablen durch Systeme diskreter 
Zahlengr\"ossen (Matrizen) ersetzt, die, von einem ganzzahligen Indexpaar 
abh\"angig, durch} algebraische \textit{Gleichungen bestimmt werden. Die Autoren 
selbst bezeichnen die Theorie als 'wahre Diskontinuumstheorie'. Die 
Undulationsmechanik hingegen bedeutet gerade umgekehrt von der 
klassischen Mechanik aus einen Schritt} auf die Kontinuumstheorie zu \textit{(...).
Im Folgenden soll nun der sehr intime} innere Zusammenhang \textit{der 
Heisenbergschen Quantenmechanik und meiner Undulationsmechanik 
aufgedeckt werden. Vom formal mathematischen Standpunkt hat man ihn 
wohl als} Identit\"at \textit{(der beiden Theorien) zu bezeichnen.''}\\*

Zur Durchf\"uhrung brauche ich nicht viel zu sagen, da wir alle damit vertraut sind, dass die Quantenmechanik in ihrer abstrakten Formulierung durch Dirac, Weyl und von Neumann die verschiedensten isomorphen Darstellungen gestattet. Dies hat die sog. Transformationstheorie zum Gegenstand.

Schr\"odinger macht sich den von F. Riesz 1907 bewiesenen mathematischen 
Sachverhalt [22] zu Nutze, dass jedes vollst\"andige Orthonormalsystem $\{u_n\}$ des $L^2$-Raumes \"uber dem Konfigurationsraum einen Hilbertraum-Isomorphismus auf den $l^2$-Raum definiert\footnote{ Tats\"achlich verwendet Schr\"odinger, wie von Neumann in seinem klassischen Buch 'Mathematische Grundlagen der Quantenmechanik' [23] in Anmerkung 35 betont, eine etwas schw\"achere Aussage, welche 1906 von Hilbert bewiesen wurde [24]. Hilbert zeigte, dass der $L^2$-Raum isomorph zu einem Unterraum von $l^2$ ist.}:
\begin{equation}
L^2 \ni \psi=\sum c_n u_n \; \longrightarrow \; \{c_n\} \in l^2.
\end{equation}
Unter diesem Isomorphismus entspricht jedem (beschr\"ankten) Operator eine Matrix und diese Zuordnung respektiert alle algebraischen Operationen, insbesondere die kanonischen Vertauschungsrelationen der Orts- und Impulsoperatoren. (Schr\"odinger verlangt f\"ur die $u_n$ hinreichende Abfalleigenschaften, damit f\"ur diese unbeschr\"ankten Operatoren bei partiellen Integrationen keine Randterme auftreten.) Besonders wichtig ist, dass die Energiematrix diagonal wird, wenn f\"ur die ${u_n}$ das System der Eigenfunktionen des Hamiltonoperators verwendet wird. (Dass kontinuierliche Spektren zus\"atzliche Schwierigkeiten verursachen war Schr\"odinger bewusst.)

Da Schr\"odinger zu diesem Zeitpunkt noch nicht im Besitz der \underline{zeitabh\"angigen} 
Wellengleichung war, ist der von ihm beschriebene Anschluss an die Heisenbergsche Bewegungsgleichung f\"ur Matrizen noch etwas merkw\"urdig. Dieser gelingt erst mit einer orthonormierten Basis $\{\varphi_n(t)\}$ von L\"osungen der zeitabh\"angigen Schr\"odinger-Gleichung
\begin{equation}
i \hbar \dot{\varphi_n}=H\varphi_n, \;\;\;\;\; (\varphi_n,\varphi_m)=\delta_{nm},
\end{equation}
die sich erst in (W6) findet. Dann gilt f\"ur die Matrix
\begin{equation}
F_{nm}=(\varphi_n,F^{op}\varphi_m)
\end{equation}
eines Schr\"odingeroperators $F^{op}$ die Heisenbergsche Bewegungsgleichung
\begin{equation}
\dot{F}=\frac{i}{\hbar}[H,F].
\end{equation}
Speziell f\"ur die station\"aren Zust\"ande $\varphi_n=u_n e^{-\frac{i}{\hbar} E_n t}$ des Hamiltonoperators ist die zugeh\"orige Matrix diagonal
\begin{equation}
H_{nm}=E_n \delta_{nm}
\end{equation}
und die Bewegungsgleichung lautet
\begin{equation}
\dot{F}_{nm}=\frac{i}{\hbar}(E_n-E_m)F_{nm},
\end{equation}
mit der L\"osung
\begin{equation}
F_{nm}(t)=F_{nm}(0) \, e^{\frac{i}{\hbar}(E_n-E_m)t}.
\end{equation}

Im bereits erw\"ahnten Brief an Jordan [18 hat Pauli all dies ebenfalls g\"anzlich unabh\"angig entwickelt. Bei ihm ist auch die Zeitabh\"angigkeit der Matrizen richtig dargestellt.

Der Ausdruck '\"Aquivalenz', bzw. 'Identit\"at' der beiden Formulierungen ist etwas \"ubertrieben, schon deshalb weil in der urspr\"unglichen Formulierung der Matrizenmechanik der Begriff 'station\"arer Zustand' gar nicht vorkommt. Darauf hat insbesondere von der Waerden in einem bekannten Aufsatz hingewiesen. Das ist aber vielleicht etwas haarspalterisch. Als Ausrutscher in der Begeisterung seiner Einsicht muss man die folgende Bemerkung Schr\"odingers entschuldigen:\\*

\textit{``In diesem Zusammenhang ist aber jedenfalls folgende Erg\"anzung des oben geschilderten \"Aquivalenzbeweises von Interesse: Die \"Aquivalenz besteht \underline{wirklich}, sie besteht \underline{auch in umgekehrter Richtung}.''}\\*

Wiederum verschiebe ich Schr\"odingers Bemerkungen zur Interpretation des 
wellenmechanischen Formalismus, bis auf die folgenden Andeutungen. Im letzten 
Abschnitt der besprochenen Arbeit betont er an einer Stelle:\\*

\textit{``Eine besonders wichtige Frage, ja vielleicht die Kardinalfrage der ganzen Atomdynamik, ist bekanntlich die Frage nach der \underline{Koppelung} zwischen dem 
atom-dynamischen Geschehen und dem elektromagnetischen Feld oder dem, 
was etwa an die Stelle des letzteren zu treten hat. (...)\\*
Hier hat nun die Matrizendarstellung der Atomdynamik auf die Vermutung 
gef\"uhrt, dass in der Tat auch das elektromagnetische Feld anders, n\"amlich 
matrizenm\"assig, dargestellt werden} muss\textit{, um die Koppelung mathematisch 
formulieren zu k\"onnen. Die Undulationsmechanik zeigt, dass hiezu jedenfalls 
kein Zwang besteht, denn der mechanische Feldskalar (von mir mit $\psi$ bezeichnet) besitzt v\"ollig die Eignung, sogar in die unver\"anderten Maxwell-
Lorentzschen Gleichungen zwischen den elektromagnetischen Feldvektoren, 
als 'Quelle' derselben einzugehen; so wie umgekehrt die elektrodynamischen 
Potentiale in die Koeffizienten der Wellengleichung eingehen, welche den 
mechanischen Feldskalar bestimmt.''}\\*

Sie sehen aus diesen Bemerkungen, dass sich Schr\"odinger eine klassisch 
realistische Interpretation erhofft. Er h\"alt aber nochmals fest, dass bei 
Mehrelektronensystemen die Wellenfunktion eine Funktion auf dem Konfigurationsraum ist, was bei einer Interpretation in dieser Richtung zu Schwierigkeiten f\"uhrt.

Darauf gab es schon fr\"uh verschiedene Reaktionen. Bereits am 24. Mai bemerkt 
Pauli an Schr\"odinger in einem inhaltsreichen Brief [25]:\\*

\textit{``Ich habe \"uberhaupt die st\"arksten Zweifel an der Durchf\"uhrbarkeit einer konsequenten reinen Kontinuums-Feld-Theorie der de Broglie-Strahlung. 
Man muss wahrscheinlich doch auch wesentlich diskontinuierliche Elemente 
in die Beschreibung der Quantenph\"anomene einf\"uhren. Ich weiss, dass diese 
Ansicht bei Dir die lebhaftesten Proteste hervorrufen wird und freue mich 
deshalb schon sehr, mit Dir ausf\"uhrlich dar\"uber zu sprechen, wenn ich Ende 
Juni nach Z\"urich komme (ich habe Debye bereits zugesagt).''}\\*

Dazu ist anzumerken, dass in Z\"urich vom 21. bis 26. Juni eine Vortragsreihe 
stattfand. Anl\"asslich dieser Tagung erfuhr \"ubrigens Schr\"odinger von Pauli und Sommerfeld von Borns Vorschlag, die Wellenfunktion statistisch zu deuten.

Etwas sp\"ater sprach Pauli vom 'Z\"urcher Lokalaberglauben', ein Ausdruck der 
rasch die Runde machte. Dazu schrieb er in einem wesentlich sp\"ateren Brief im November bes\"anftigend an Schr\"odinger [26]:\\*

\textit{``Was meine Bemerkung \"uber den 'Z\"uricher Lokalaberglauben' betrifft, so m\"ochte ich Dich sehr bitten, sie nicht als pers\"onliche Unfreundlichkeit Dir gegen\"uber, sondern als Ausdruck der sachlichen \"Uberzeugung anzusehen, dass die Quantenph\"anomene in der Natur solche Seiten zeigen, die nicht mit 
den Begriffen der Kontinuumsphysik (Feldphysik) allein erfasst werden 
k\"onnen. Glaube aber} nicht\textit{, dass mir diese \"Uberzeugung das Leben leicht 
macht, ich habe mich um ihretwillen schon sehr geplagt und werde das wohl 
auch noch weiter tun m\"ussen!''}\\*

Aber kehren wir zur Schr\"odingerschen Sequenz zur\"uck.

\section*{Wellenmechanische St\"orungstheorie}
\"Uber die dritte Mitteilung der Serie 'Quantisierung als Eigenwertproblem' [1] brauche ich nicht viel zu sagen. In dieser langen Arbeit von 53 Seiten entwickelt Schr\"odinger die station\"are St\"orungstheorie und berechnet als Anwendung die Aufspaltung beim Starkeffekt. F\"ur den anomalen Zeeman-Effekt ist es noch zu fr\"uh. Dazu bemerkt er:

\textit{``Man wird versuchen m\"ussen, den Uhlenbeck-Goudsmitschen Gedanken in die Wellenmechanik aufzunehmen.''}

Dies hat dann, wie Sie wissen, einige Zeit sp\"ater Pauli bewerkstelligt.

\section*{Die zeitabh\"angige Wellengleichung}
Ich komme zu Schr\"odingers letzter Arbeit der wellenmechanischen Serie. Endlich findet er auch die \underline{zeitabh\"angige} Gleichung. Sie werden sich wahrscheinlich auch wundern, weshalb Schr\"odinger damit so viel M\"uhe hatte. Der Hauptgrund beruht darauf, dass er f\"ur eine realistische wellenmechanische Atomtheorie eine \underline{reelle} Wellenfunktion bevorzugte. Ich belege dies mit dem Schlussabschnitt der langen Arbeit:\\*

\textit{``Eine gewisse H\"arte liegt ohne Zweifel zurzeit noch in der Verwendung einer} komplexen \textit{Wellenfunktion. W\"urde sie} grunds\"atzlich \textit{unvermeidlich und nicht eine blosse Rechenerleichterung sein, so w\"urde das heissen, dass grunds\"atzlich} zwei \textit{Wellenfunktionen existieren, die erst} zusammen \textit{Aufschluss \"uber den Zustand des Systems geben. Diese etwas unsympathische Folgerung l\"asst, wie ich glaube, die sehr viel sympathischere Deutung zu, dass der Zustand des Systems durch eine reelle Funktion und 
ihre Ableitung nach der Zeit gegeben ist. Dass wir hier\"uber noch keinen 
genaueren Aufschluss geben k\"onnen, h\"angt damit zusammen, dass wir in dem 
Gleichungspaar (4'') nur das - f\"ur die Rechnung allerdings ausserordentlich 
bequeme -} Surrogat \textit{einer reellen Wellengleichung von wahrscheinlich der vierten Ordnung vor uns haben, deren Aufstellung mir jedoch im nichtkonservativen Fall nicht gelingen wollte.''}

Diese letzte Aussage muss ich n\"aher erkl\"aren. In \S1 betrachtet Schr\"odinger zun\"achst \underline{konservative} Systeme. F\"ur solche gen\"ugen nach seinen vorangegangen Arbeiten die station\"aren Zust\"ande der Eigenwertgleichung $Hu = Eu$ und damit der Gleichung 4. Ordnung:
\begin{equation}
H^2 u=E^2 u.
\end{equation}
Betrachten wir mit Schr\"odinger eine \underline{reelle} Wellenfunktion mit harmonischer 
Zeitabh\"angigkeit
\begin{equation}
\psi=\Re(u e^{\pm i \omega t}), \;\;\;\;\; \omega=E/\hbar
\end{equation}
so ist die Multiplikation mit $E^2$ \"aquivalent zu $-\hbar^2 \partial^2/\partial t^2$ und Schr\"odinger erh\"alt anstelle von (37) die Gleichung
\begin{equation}
H^2 \psi=-\hbar^2 \frac{\partial^2 \psi}{\partial t^2}.
\end{equation}
Dazu sagt er: \textit{``[Diese] Gleichung ist daher offenbar die \underline{einheitliche} und \underline{allgemei-}\\
\underline{ne} Wellengleichung f\"ur den Feldskalar $\psi$''}, und er betont, dass diese von 4. Ordnung -wie bei vielen Problemen der Elastizit\"atstheorie ( z. B. schwingende Platte)- ist.

Die \"Ubertragung von (39) auf \underline{nichtkonservative} Systeme ist aber nicht akzeptabel, denn dann m\"usste ein Glied mit $\partial V/ \partial t$ vorkommen. F\"ur solche Systeme betrachtet Schr\"odinger zun\"achst doch \underline{komplexe} Wellenfunktionen. L\"asst man in (38) den Realteil weg, so ist die Multiplikation mit $E$ 
\"aquivalent zu $\pm i \hbar \partial/\partial t$, weshalb Schr\"odinger zum Paar
\begin{equation}
H \psi = \pm i \hbar\frac{\partial \psi}{\partial t}
\end{equation}
gef\"uhrt wird. Dazu sagt er:\\*

\textit{``Wir werden verlangen, dass die komplexe Wellenfunktion $\psi$ einer dieser beiden Gleichungen gen\"uge. Da alsdann die konjugiert komplexe Funktion $\overline{\psi}$ der anderen Gleichung gen\"ugt, wird man als reelle Wellenfunktion (wenn man sie ben\"otigt) den Realteil von $\psi$ ansehen d\"urfen.''}\\*

Im Besitz dieser Grundgleichung entwickelt Schr\"odinger nun die \underline{zeitabh\"angige} St\"orungstheorie und wendet diese auf die Dispersionstheorie an. Er berechnet dabei st\"orungstheoretisch das induzierte Dipolmoment eines Atoms in einem zeitabh\"angigen Feld und begr\"undet so die Formeln von Kramers und Heisenberg, welche 1924 auf der Basis von korrespondenzm\"assigen Betrachtungen gewonnen wurden. (Auch \"uber den Resonanzfall l\"asst er sich lange aus, ohne eine L\"osung anzubieten; dies wurde erst viel sp\"ater durch Weisskopf und Wigner geleistet.)

In einem sp\"ateren Abschnitt (\S6) erh\"alt Schr\"odinger auf sehr formalem Weg die \textit{'relativistisch-magnetische Verallgemeinerung des Feldskalars'.}

Der letzte \S7 hat die \"Uberschrift: \textit{'\"Uber die physikalische Bedeutung des 
Feldskalars'.} Dies f\"uhrt uns zu den Interpretationsfragen der neuen, mathematisch ausformulierten Theorie.

\section*{Auseinandersetzungen zu Interpretation}
In dem erw\"ahnten Schlussparagraphen deutet Schr\"odinger $\overline{\psi}\psi$ als 'eine Art \underline{Gewichtsfunktion} im Konfigurationsraum' und zeigt, dass f\"ur diese die bekannte Kontinuit\"atsgleichung im Konfigurationsraum gilt. Im Anschluss daran macht er ein paar vage Andeutungen im Sinne einer realistischen elektromagnetischen Interpretation, die er mit den Worten beschliesst:\\*

\textit{``Ich hoffe und glaube, dass die vorstehenden Ans\"atze sich zur Erkl\"arung der magnetischen Eigenschaften der Atome und Molek\"ule und weiterhin auch zur Erkl\"arung der Elektrizit\"atsstr\"omung in festen K\"orpern als n\"utzlich erweisen werden.''}\\*

W\"ahrend der ganze Formalismus der Wellenmechanik von allen dankbar aufgenommen wurde, stiess Schr\"odinger bei einem Grossteil der massgebenden Kollegen 
hinsichtlich seinen physikalischen Anschauungen auf Widerstand. Einiges dazu habe ich bereits zitiert. In einem Brief von 8. Juni an Pauli \"aussert Heisenberg seinen ganzen Unmut [27]:\\*

\textit{``Je mehr ich \"uber den physikalischen Teil der Schr\"odingerschen Theorie nachdenke, desto abscheulicher finde ich ihn. Was Schr\"odinger \"uber 
Anschaulichkeit seiner Theorie schreibt ... ich finde es Mist.''}\\*

Nachdem Heisenberg in einer ber\"uhmten Arbeit die Wellenmechanik auf das 
Heliumatom angewandt hatte, schreibt er am 26. Juli wiederum an Pauli [28]:\\*

\textit{``So nett Schr\"odinger pers\"onlich ist, so merkw\"urdig find' ich seine Physik: man kommt sich, wenn man sie h\"ort, um 26 Jahre j\"unger vor. Schr\"odinger wirft ja alles 'quantentheoretische': n\"amlich lichtelektrischen Effekt, Francksche St\"osse, Stern-Gerlacheffekt usw. \"uber Bord (...).''}\\*

Es gab nat\"urlich auch andere Stimmen, vor allem unter den Altmeistern, wie etwa W. Wien und zun\"achst auch Einstein. Freilich kamen bei letzerem sehr bald Zweifel auf.

Zur h\"artesten Auseinandersetzung kam es dann in Kopenhagen Ende September im Hause von N. Bohr. (Auch Heisenberg und Pauli befanden sich damals in Kopenhagen.) Heisenberg hat \"uber diese intensiven Diskussionen in 'Der Teil und das Ganze'' berichtet [29]. Hier ein paar S\"atze aus diesem Bericht:\\*

\textit{``Die Diskussionen zwischen Bohr und Schr\"odinger begannen schon auf dem Bahnhof von Kopenhagen und wurden jeden Tag vom fr\"uhen Morgen bis sp\"at 
in die Nacht hinein fortgesetzt. Schr\"odinger wohnte bei Bohrs im Hause, so 
dass es schon aus \"ausseren Gr\"unden kaum eine Unterbrechung der Gespr\"ache 
geben konnte. Und obwohl Bohr sonst im Umgang mit Menschen besonders 
r\"ucksichtsvoll und liebensw\"urdig war, kam er mir hier beinahe wie ein 
unerbittlicher Fanatiker vor, der nicht bereit war, seinem Gespr\"achspartner 
auch nur einen Schritt entgegenzukommen oder auch nur die geringste 
Unklarheit zuzulassen. Es wird kaum m\"oglich sein wiederzugeben, wie 
leidenschaftlich die Diskussionen von beiden Seiten gef\"uhrt wurden, wie tief 
verwurzelt die \"Uberzeugungen waren, die man gleichermassen bei Bohr und 
Schr\"odinger hinter den ausgesprochenen S\"atzen sp\"uren konnte. So kann es 
sich im folgenden nur um ein sehr blasses Abbild jener Gespr\"ache handeln, in 
denen mit \"ausserster Kraft um die Deutung der neugewonnenen 
mathematischen Darstellung der Natur gerungen wurde. (...)\\*
So ging die Diskussion \"uber viele Stunden des Tages und der Nacht, ohne 
dass es zu einer Einigung gekommen w\"are. Nach einigen Tagen wurde 
Schr\"odinger krank, vielleicht als Folge der enormen Anstrengung; er musste 
mit einer fiebrigen Erk\"altung das Bett h\"uten. Frau Bohr pflegte ihn und brachte Tee und Kuchen, aber Niels Bohr sass auf der Bettkante und sprach auf 
Schr\"odinger ein: $\ll$Aber Sie m\"ussen doch einsehen, dass ...$\gg$''}\\*

Schr\"odinger erkannte mit der Zeit schon, dass er seine urspr\"unglichen 
Vorstellungen nicht aufrecht erhalten konnte. Die folgende Passage stammt aus einem Brief von ihm an Sommerfeld vom 29. April 1927 [30].\\*

\textit{``Bez\"uglich der Deutung der Quantenmechanik bin ich unsicherer denn je. Die Arbeiten Uns\"olds und Paulings geben mir zwar wieder grosses Vertrauen zu den 'verschmierten Elektronen' aber die Schwierigkeiten der 
Kontinuumsauffassung lassen sich doch nicht fortleugnen. Ich kann sogar 
schon ein Bisschen das \"Argernis verstehen, das ich vielen gegeben habe 
dadurch, dass ich in einer grossen, wenig kritischen Menge ein Siegesgeheul 
wachgerufen habe: 'Nieder mit den Quanten! Die Kontinuumsauffassung 
gerettet!' Ich muss den anderen ein Bisschen als Demagoge erscheinen, der 
auf die Leichtgl\"aubigkeit der Menge spekuliert und ihr nach Wunsch redet. - 
Nun, es wird sich schon kl\"aren. Der st\"urmende Most, den wir jetzt von allen Seiten schl\"urfen, gibt [?] noch einen ganz guten Wein.''}\\*

Mit der 'Kopenhagen-Deutung' der Quantenmechanik konnte sich Schr\"odinger 
jedoch nicht anfreunden. Sehr \"ahnlich wie Einstein war es ihm unm\"oglich, den Realit\"atsbegriff aufzugeben, welcher der klassischen Physik zugrunde liegt. Ich bin aber der Meinung, dass die scharfe Kritik der beiden auch positive Auswirkungen hatte. Mit ihren Analysen und Gedankenexperimenten belebten sie die Diskussion um die Interpretationsfragen bis auf den heutigen Tag ganz wesentlich. Schr\"odingers 'Verschr\"anktheit der Zust\"ande' und sein 'Katzen-Experiment' sind noch immer in aller Munde und haben schliesslich auch zu tats\"achlichen Experimenten gef\"uhrt, die vermutlich in der weiteren Zukunft sogar wichtige technologische Auswirkungen nach sich ziehen werden. Dar\"uber werden andere an diesem Symposium berichten.

Geblieben ist Schr\"odingers mathematisch ausformulierte Theorie der 
Wellenmechanik, insbesondere seine fundamentale dynamische Gleichung. An ihr hat sich nichts ge\"andert und sie wird solange unterrichtet werden als Menschen 
Naturwissenschaft betreiben.

\section*{Literaturangaben}

\begin{enumerate}
\item Die quantentheoretischen Beitr\"age von Schr\"odinger finden sich alle in Band 3 der Gesammelten Abhandlungen, Ref. (2).\\*
     
Die Originalzitate der sechs Arbeiten zur Wellenmechanik, in folgendem mit (W1)-(W6) abgekürzt, sind:\\*
\begin{itemize}
\item(W1) E. Schr\"odinger, \textit{Quantisierung als Eigenwertproblem}\\*
 (Erste Mitteilung), Ann. Phys. \textbf{79}, 361-76 (1926).\\*

\item(W2) E. Schr\"odinger, \textit{Quantisierung als Eigenwertproblem}\\*
 (Zweite Mitteilung), Ann. Phys. \textbf{79}, 489-527 (1926).\\*

\item(W3) E. Schr\"odinger, \textit{Der stetige \"Ubergang von der Mikro-\\*
 zur Makromechanik},\\*
 Die Naturwissenschaften, 14. Jahrg. Heft 28, S. 664-666 (1926).\\*

\item(W4) E. Schr\"odinger, \textit{\"Uber das Verh\"altnis der Heisenberg-Born-Jordanschen\\*
 Quantenmechanik zu der meinen}, Ann. Phys. \textbf{79}, 734-56 (1926).\\*

\item(W5) E. Schr\"odinger, \textit{Quantisierung als Eigenwertproblem}\\*
 (Dritte Mitteilung), Ann. Phys. \textbf{80}, 437-90 (1926).\\*

\item(W6) E. Schr\"odinger, \textit{Quantisierung als Eigenwertproblem}\\*
 (Vierte Mitteilung), Ann. Phys. \textbf{81}, 109-39 (1926).\\*
\end{itemize}

\item E. Schr\"odinger, Gesammelte Abhandlungen. Verlag der \"Osterreichischen Akademie der Wissenschaften, Friedr. Vieweg \& Sohn, Braunschweig/Wiesbaden, Wien 1984, 4 Bde.\\*

\item K. von Meyenn, Schr\"odinger-Korrespondenz, in Vorbereitung.\\*

\item E. Schr\"odinger, Brief an W. Wien, 22. Feb. (1926), Reg. (3).\\*

\item A. Pais, Albert Einstein, Eine wissenschaftliche Biographie, Vieweg (1986).\\*

\item A. Einstein, Brief an H.A. Lorentz, 16. Dez. (1924).\\*

\item A. Einstein, Sitzungsb. Preuss. Akad. Wiss. 1925, S. 3.\\*

\item A. Einstein, Phys. Zeitschr. \textbf{10}, 185 (1909); \textbf{10}, 817 (1909).\\*

\item W. Pauli, \textit{Einsteins Beitrag zur Quantentheorie}, in Physik und Erkenntnistheorie, Vieweg, Braunschweig (1984).\\*

\item E. Schr\"odinger, Phys. Z. \textbf{27}, 95-101(1926).\\*

\item F. Bloch, Physics today \textbf{29}, (December),  23-7 (1976).\\*

\item W. Moore, \textit{Schr\"odinger, Life and Thought}, Cambridge University Press (1989), p. 195.\\*

\item E. Schr\"odinger, Brief an W. Wien, 27. Dez. (1925), Ref. (3).\\*

\item E. Schr\"odinger, Brief an W. Yourgrau, Jan. 1956, Ref. (12), p. 196.\\*

\item E. Schr\"odinger, Brief an A. Sommerfeld, 29. Jan. (1926), Ref. (3).\\*

\item A. Sommerfeld, Brief an E. Schr\"odinger, 3. Feb. (1926), Ref. (3).\\*

\item A. Sommerfeld, Brief an W. Pauli, 3. feb. (1926), in \textit{Wissenschafticher Briefwechsel mit Bohr, Einstein, Heisenberg u.a.}, Band I: l919-1929. Herausgegeben von A. Hermann, K. v. Meyenn und V. F. Weisskopf. Springer, New York/Heidelberg/Berlin 1979. Brief [118a], in Band II abgedruckt.\\*

\item W. Pauli, Brief an P. Jordan, 12. April (1926), Ref. (17), Dok. [131].\\*

\item Siehe, z.b., V. I. Arnold, Mathematical Methods of Classical Mechanics, Graduate Texts in Mathematics 60, Springer-Verlag, New York/Heidelberg/Berlin (1978); speziell §45.\\*

\item H. A. Lorentz, Brief an E. Sommerfeld, 27. Mai (1926), in \textit{Briefe zur Wellenmechanik}, herausgegeben von K. Przibram, Springer-Verlag, Wien (1963), p. 41\\*

\item H. A. Lorentz, Brief an E. Schr\"odinger, 19. Juni (1926), Ref. (20), p. 61.\\*

\item F. Riesz, C. R. Acad. Sci., Paris \textbf{144}, 615-619 (1907).\\*

\item J. von Neumann, \textit{Mathematische Grundlagen der Quantenmechanik}, Springer-Verlag, Berlin/Heidelberg (1996).\\*

\item D. Hilbert, G\"ott. Nachr. (1906).\\*

\item W. Pauli, Brief an E. Schr\"odinger, 24. Mai (1926), Ref. (17), Dok. [134].\\*

\item W. Pauli, Brief an E. Schr\"odinger, 22. Nov. (1926), Ref. (17), Dok. [147].\\*

\item W. Heisenberg, Brief an W. Pauli, 8. Juni (1926), Ref. (17), Dok. [136].\\*

\item W. Heisenberg, Brief an W. Pauli, 28. Juli (1926), Ref. (17), Dok. [142].\\*

\item W. Heisenberg, \textit{Der Teil und das Ganze}, Piper-Verlag, M\"unchen (1969); Kap. 6.\\*

\item E. Schr\"odinger, Brief an A. Sommerfeld, 29. April (1927), Ref. (3).\\*

\end{enumerate}
\end{document}